# Viscoelastic Behavior of Rubbery Materials

C.M. Roland

*Polymer Physics Section*
*Naval Research Laboratory*
*Washington, DC*

## OXFORD
### UNIVERSITY PRESS

# Contents

# CHAPTER 4
# Networks

Rubber can be deformed to a substantial degree, which gives rise to deviations from stress-strain linearity. An elastic nonlinear response (strain energy depends only on the initial and final states) is referred to as hyperelasticity. There are two approaches to analyzing the large strain behavior of rubbery networks, phenomenological and based on statistical or molecular models. The former describe the deformation using strain energy functions drawn from continuum mechanics, with simplifying restrictions yielding tractable constitutive relations. Molecular models use statistical mechanics to compute the entropy from consideration of the topology of the chains and its modification by strain; the intramolecular energy contribution to the elastic energy (e.g., the energy difference of gauche and trans backbone bonds) may also be included in the analysis. Early rubber elasticity models[*] ignored intermolecular interactions, whereas later developments included the effect of entanglements or steric constraints on the mechanical stress.

Although the focus of rubber elasticity theories is the chain configurations, the segmental and chain dynamics are governed by the same intramolecular and intermolecular potentials and correlations. Thus, insights into the viscoelastic response can be gleaned from analysis of equilibrium mechanical properties. These parallels between dynamics and the equilibrium behavior are brought out by the coupling model of relaxation [1,2], introduced in Chapter 2. For this reason and because of its focus on experimental results, we use the model to interpret the relaxation properties in terms of rubber elasticity concepts.

Since network properties *per se* involve primarily long length scales, they depend only weakly on the chemical structure. Hence, all elastomers exhibit essentially the same mechanical behavior up through moderate strains, whatever polymer comprises the network. This axiom implies that attempts to improve the mechanical properties will have limited success. Failure properties such as strength and fatigue life exhibit a maximum versus crosslink density **Figure 2** [3]. In order to minimize creep, the practically useful range of crosslinking falls past this maximum; thus, elastomer failure properties vary inversely with the stiffness: increasing crosslink densities yield higher modulus but lower strength and fatigue life. A given rubber compound represents a compromise between stiffness and strength; both properties cannot be optimized using conventional approaches. To circumvent this "fundamental" limitation requires unorthodox structures or morphologies. Five examples of alternative networks – interpenetrating, double, bimodal, heterogeneous, and deswollen – are discussed at the conclusion of the chapter.

---

[*] Early molecular theories of rubber hyperelasticity were called kinetic theories, in correspondence to the behavior of ideal gases (the molecules of which have neither size nor mutual interactions). The elastic forces of an ideal rubber are caused by thermal motion of the network chains attempting to restore the higher entropy, isotropic state; thus, analogous to ideal gases, ideal rubber elasticity is a statistical effect. In real materials there is also an energy contribution (negative or positive) due to the difference in energy between trans and gauche conformations of the repeat units.

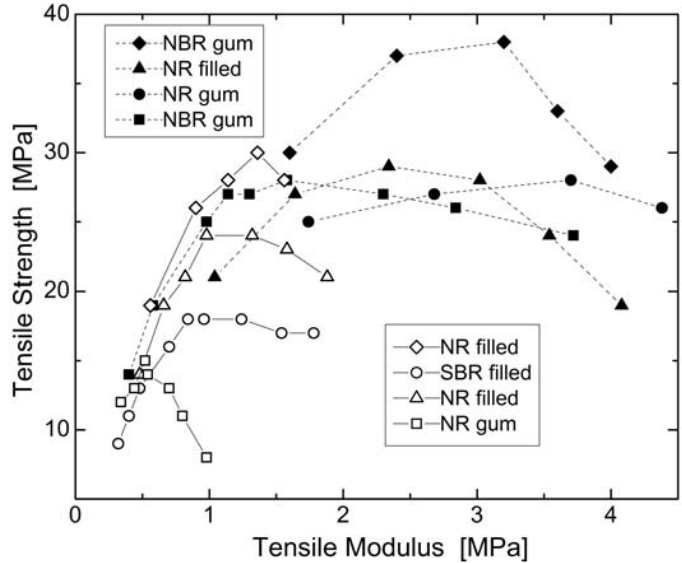

Figure 1. Tensile strength *versus* modulus for elastomers with covalent (hollow symbols) and ionic (solid symbols) crosslinks. Failure properties go through a maximum as a function of the degree of crosslinking. Data from ref. [3].

## 4.1. Phenomenological elasticity models

Phenomenological treatments [4,5,6] use mathematics to describe the mechanical potential without attempting to explain its molecular origin. The relevant potential depends on the experimental conditions: Gibbs free energy for isothermal, isobaric processes; Helmholtz free energy for isothermal, isochoric conditions; and enthalpic energy for adiabatic, isochoric conditions. The deformation is assumed to be reversible, so only the elastic strain energy, $W$, is derived, with the principal stresses[†] obtained as the partial derivatives of $W$ for the relevant conditions (constant $T$, $P$, etc.). A further simplification is that the strain is uniform, without localized phenomenon such as necking. The analysis is often carried out in terms of the strain invariants, which are scalar combinations of the strain components, rather than the strains themselves. Strain invariants are independent of the axes used to define the sample geometry, enabling calculations for inhomogeneous deformations without explicit consideration of the principal directions (during a homogenous deformation, parallel lines remain parallel). Rotation of the body corresponds to a change in sign of two of the three principal stretch ratios, and in order for the strain energy to be independent of rotation, the strain invariants must be even functions of the strain. The simplest invariants are then symmetric functions of the squares of the stretch ratios[‡]

$$I_1 = \lambda_1^2 + \lambda_2^2 + \lambda_3^2 \ , \ \ I_2 = \lambda_1^2\lambda_2^2 + \lambda_1^2\lambda_3^2 + \lambda_2^2\lambda_3^2 \ , \ \ I_3 = \lambda_1^2\lambda_2^2\lambda_3^2 \tag{5.1}$$

where the subscripts denote one of the three rectilinear coordinates. The elastic strain energy depends only on the current strain and since the body is isotropic, $W$ is assumed to be symmetrical with the strain components; thus,

$$W = f(I_1, I_2, I_3) \tag{5.2}$$

---

[†] Principal stresses are aligned with the coordinate system such that cross-terms are zero; that is, the stress tensor is a diagonal matrix.

The strain energy of any isotropic material can be expressed as a function of these three strain invariants. Since the bulk modulus of rubber is typically 3 orders of magnitude larger than the shear modulus, the assumption of incompressibility entails negligible error; hence, $I_3$ is taken to be unity, reducing eqn (5.2) to

$$W = f(I_1, I_2) \tag{5.3}$$

The corresponding principal stress differences are given by the derivatives of $W$ [7]

$$\sigma_1\lambda_1 - \sigma_2\lambda_2 = 2\left(\lambda_1^2 - \lambda_2^2\right)\left(\frac{\partial W}{\partial I_1} + \lambda_3^2\frac{\partial W}{\partial I_2}\right) \tag{5.4}$$

$$\sigma_1\lambda_1 - \sigma_3\lambda_3 = 2\left(\lambda_1^2 - \lambda_3^2\right)\left(\frac{\partial W}{\partial I_1} + \lambda_2^2\frac{\partial W}{\partial I_2}\right) \tag{5.5}$$

and

$$\sigma_2\lambda_2 - \sigma_3\lambda_3 = 2\left(\lambda_2^2 - \lambda_3^2\right)\left(\frac{\partial W}{\partial I_1} + \lambda_1^2\frac{\partial W}{\partial I_2}\right) \tag{5.6}$$

Or in terms of the true stresses and derivatives with respect to the stretch ratios

$$\hat{\sigma}_1 - \hat{\sigma}_2 = \lambda_1\frac{\partial W}{\partial \lambda_1} - \lambda_2\frac{\partial W}{\partial \lambda_2} \ , \ ... \tag{5.7}$$

From these equations experimental stresses can be calculated for any given mode of strain. The relevant expressions are listed in Table 1 for the primary strain modes. The problem is to identify the form of the function $f$ in eqn (5.3).

---

[‡] There are various measures of strain, or relative displacement of points in a material. For normal strains the stretch (or extension) ratio, $\lambda$, which is the ratio of the final and initial dimensions parallel to the load, equals the engineering (or Cauchy) strain plus one. The true (or Hencky) strain is the natural logarithm of the stretch ratio, and is equal to the engineering strain for infinitesimal increments. Confusingly, while the Cauchy strain is the engineering strain, the Cauchy stress is the true stress, not the engineering stress. Normal strains act perpendicular to a body to stretch or compress it, whereas shear strains act parallel to a surface, inducing distortion.

Table 1. Stretch-stress relations for an incompressible material.

| Deformation mode | Principal stretch ratios | Engineering Stress | Equivalent strain |
|---|---|---|---|
| Simple shear | na | $\sigma = 2\left(\dfrac{\partial W}{\partial I_1} + \dfrac{\partial W}{\partial I_2}\right)(\lambda - \lambda^{-1})$ | Tension plus rotation |
| Pure shear | $\lambda_1 = \lambda$ $\lambda_2 = 1$ $\lambda_3 = \lambda^{-1}$ | $\sigma_1 = 2\left(\dfrac{\partial W}{\partial I_1} + \dfrac{\partial W}{\partial I_2}\right)(\lambda - \lambda^{-3})$ | Plane strain compression |
| Uniaxial strain | $\lambda_1 = \lambda$ $\lambda_2 = \lambda_3 = \lambda^{-\frac{1}{2}}$ | $\sigma = 2\left(\dfrac{\partial W}{\partial I_1} + \lambda^{-1}\dfrac{\partial W}{\partial I_2}\right)(\lambda - \lambda^{-2})$ | Tension or compression |
| Biaxial strain | for equal biaxial: $\lambda_1 = \lambda_2 = \lambda$ $\lambda_3 = \lambda^{-\frac{1}{2}}$ | $\sigma_1 = 2\left(\dfrac{\partial W}{\partial I_1} + \dfrac{\partial W}{\partial I_2}\lambda_2^2\right)\left(\lambda_1 - \lambda_1^3\lambda_2^2\right)$ $\sigma_2 = 2\left(\dfrac{\partial W}{\partial I_1} + \dfrac{\partial W}{\partial I_2}\lambda_1^2\right)\left(\lambda_2 - \lambda_1^2\lambda_2^3\right)$ | Inflation |

A general expression for the strain energy that goes to zero in the absence of strain is

$$W = \sum_{i=0, j=0}^{\infty} C_{ij}(I_1 - 3)^i (I_2 - 3)^j \qquad \left(C_{00} = 0\right) \qquad (5.8)$$

The first term in this series, $i = 1$ and $j = 0$, is identical to the classical rubber elasticity result (eqn (5.32) below). The other linear term, $i = 0$ and $j = 1$, cannot alone describe actual data. Summing both these terms yields the well-known Mooney-Rivlin equation [4,5]

$$W = C_{10}(I_1 - 3) + C_{01}(I_2 - 3) \qquad (5.9)$$

The linearity in the invariants simplifies application to experimental measurements. For uniaxial strain the engineering stress along the stretch direction becomes

$$\sigma = (C_{10} - C_{01} / \lambda)(\lambda - \lambda^{-2}) \qquad (5.10)$$

the other two principal stresses being zero. For shear strains the Mooney-Rivlin stress is

$$\sigma = 2(C_{10} - C_{01})\gamma \qquad (5.11)$$

In **Figure 3** data measured for a NR network subjected to tension and compression are plotted in the Mooney-Rivlin form of the reduced stress, $\sigma / (\lambda - \lambda^{-2})$, versus the reciprocal of the stretch ratio. Eqn (5.10) predicts a straight line; however, even neglecting the upturn at high extensions, the slopes for tension and compression differ significantly. This is well known − different modes of deformation yield different values of the Mooney-Rivlin elastic constants; thus, fitting eqn (5.9) to the tension data yields a different set of elastic constants than fitting to the compression results. An expedient is to set $C_{01} = 0$ for shear while using it as an adjustable parameter for the tensile data. While this improves the fit, it lacks any basis. The Mooney-Rivlin equation is not a valid constitutive equation and cannot be used to predict stresses for other modes from measurements in a given mode (usually tension). The Mooney-Rivlin equation also fails at higher strains, at which there is an upturn in the stress as the network chains approach their ultimate deformability (due to non-Gaussian effects discussed below). This upturn is seen **Figure 3**.

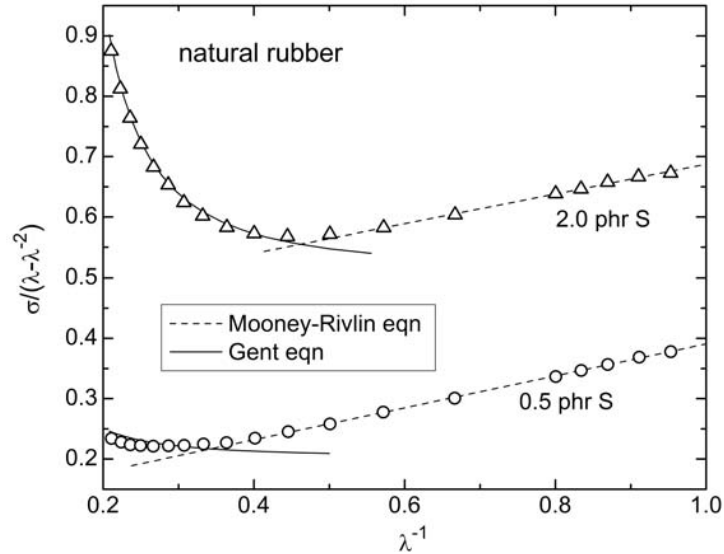

Figure 2. Mooney-Rivlin plot of reduced engineering stress versus reciprocal stretch ratio for NR networks with two degrees of crosslinking. The dashed line is a linear fit to the higher strains; that is, eqn (5.10) with $C_{10}$ = .063 and .132 MPa for 0.5 and 2.0 phr sulfur levels, respectively. The solid line is the fit to eqn (5.26) using $C_{10}$ = 0.103 MPa and $I_m$ = 128 for the lower crosslink density, and $C_{10}$ = 0.263 MPa and $I_m$ = 52 for the higher. Data from ref. [25].

An obvious extension, found in finite analysis software, is the addition of more terms in eqn (5.8). Two popular forms for fitting experimental results for various modes of strain are a three term expression [8]

$$W = C_{10}(I_1 - 3) + C_{01}(I_2 - 3) - C_{02}(I_2 - 3)^2 \qquad (5.12)$$

or the higher order cubic equation [9,10]

$$W = C_{10}(I_1 - 3) + C_{01}(I_2 - 3) + C_{20}(I_1 - 3)^2 + C_{11}(I_1 - 3)(I_2 - 3) + C_{30}(I_1 - 3)^3 \qquad (5.13)$$

Ref. [11] is an extensive bibliography of various efforts to approximate eqn (5.8) to obtain useful but not overly cumbersome expressions for the strain energy. While higher order terms improve the accuracy of fits to experimental data, their extrapolation can lead to larger errors than the simpler Mooney-Rivlin equation.

Ogden [12] proposed an alternative form in terms of the stretch ratios rather than the invariants, which also included the odd terms omitted in the Mooney-Rivlin formulation

$$W = \sum_i^M \frac{a_i}{b_i} (\lambda_1^{b_i} + \lambda_2^{b_i} + \lambda_3^{b_i} - 3) \tag{5.14}$$

The exponents $b_i$ are not necessarily integers or even positive; rather, their values are chosen to best describe the mechanical response of a particular material. The Ogden equation simplifies the analysis since, unlike eqn.(5.8), all terms have the same form. From eqn.(5.7) the principal stress differences are

$$\hat{\sigma}_1 - \hat{\sigma}_2 = \sum_i^M a_i (\lambda_1^{b_i} - \lambda_2^{b_i}) \tag{5.15}$$

which for uniaxial strain gives

$$\sigma = \sum_i^M a_i (\lambda^{b_i - 1} - \lambda^{-\frac{b_i}{2} - 1}) \tag{5.16}$$

and for shear strain

$$\sigma = \sum_i^M a_i (\lambda^{b_i - 1} - \lambda^{-b_i - 1}) \tag{5.17}$$

The number of terms, $M$, depends on the span of the experimental data and the desired accuracy of the fitting. For a general description, three terms involving six adjustable parameters are usually necessary, although for finite element modeling involving only two modes, such as axial and torsional deformations, $M = 2$ may suffice [12,13,14].

There are numerous variations on eqn (5.8) that invoke *ad hoc* forms for the functional dependence of the strain energy. Valanis and Landel [15], extending the earlier work of Carmichael and Holdaway [16], expressed $W$ as the sum of three terms, one for each of the three principal strains, all having the same functional form

$$W = w(\lambda_1) + w(\lambda_2) + w(\lambda_3) \tag{5.18}$$

This separability is consistent with statistical analyses of Gaussian chains that assume the fluctuations in orthogonal directions are independent. Different functions have been used for the $w$ [17,18], and with certain restrictions eqn (5.3) can be expressed in the Valanis-Landel form [8,19]. Derivatives of eqn (5.18) yield the principal stresses

$$\hat{\sigma}_1 - \hat{\sigma}_2 = \lambda_1 \frac{dw}{d\lambda_1} - \lambda_2 \frac{dw}{d\lambda_2} \ , \ ... \tag{5.19}$$

which makes clear the advantage of the separability of the Valanis-Landel form. For example, if experiments are carried out at constant $\lambda_2$, such as pure shear ($\lambda_1 = \lambda_3^{-1}$; $\lambda_2 = 0$), the variation of $w$ with $\lambda_1$ is obtained; thus, specification of the Valanis-Landel equation requires only measurements of a single function of just one variable.

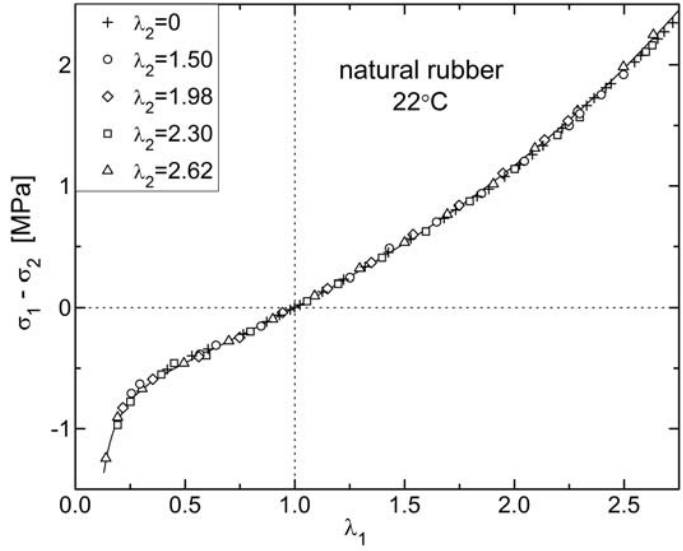

Figure 3. Stresses measured for an NR network as a function of $\lambda_1$. The biaxial strain data for each value of $\lambda_2$ were superposed onto the pure shear results ($\lambda_2 = 0$) by addition of a constant. The solid line is the fit of the Ogden equation (eqn (5.20)).

**Figure 4** shows stresses for a NR network measured at various $\lambda_1$ for different, fixed constant values of $\lambda_2$; that is, for pure shear and various biaxial deformations ($\sigma_1 \geq \sigma_2$; $\sigma_3 = 0$). According to eqn (5.19), the principal stresses for each strain mode differ only by a constant, and accordingly, when the quantity $\left(dw/d\lambda_2\right)\lambda_2$ is added to the experimental stresses, the data collapse to a single curve. Since the Ogden equation has the Valanis-Landel form, it can describe the data in **Figure 4**, as shown by the fit of eqn (5.15) using

$$\lambda\frac{dw}{d\lambda} - c_{\lambda_2} = 0.69(\lambda^{1.3}-1) + 0.01(\lambda^{4.0}-1) - 0.0122(\lambda^{-2.0}-1) \qquad (5.20)$$

with the constant $c_{\lambda_2}$, reflecting the contribution of the $\lambda_2$ term, adjusted to superpose the curves.

The above constitutive relations poorly describe stresses over the entire range from low strains up through deformations at which finite extensibility of the chains becomes significant [20]. Yeoh [21] accounted for the high strain elastic behavior by assuming that $\partial W/\partial I_2 = 0$. While not strictly correct, $\partial W/\partial I_2$ is small for filled rubbers [22,23] and the upturn in the stress occurs at lower strains than in the absence of filler. With this assumption eqn (5.13) becomes [21]

$$W = C_{10}(I_1-3) + C_{20}(I_1-3)^2 + C_{30}(I_1-3)^3 \qquad (5.21)$$

The strain energy thus depends only on the value of $I_1$ for any particular mode of strain. The uniaxial stress is

$$\sigma = \left[2C_{10} + 4C_{20}(I_1-3) + 6C_{30}(I_1-3)^2\right]\left(\lambda - \lambda^{-2}\right) \qquad (5.22)$$

and the shear stress

$$\sigma = \left[ 2C_{10} + 4C_{20}(I_1 - 3) + 6C_{30}(I_1 - 3)^2 \right] \gamma \qquad (5.23)$$

Unlike the Mooney-Rivlin equation, the shear modulus, $G = \sigma/\gamma$, in the Yeoh model varies with strain. **Figure 5** displays the uniaxial tension data from **Figure 3** plotted versus the strain invariant $I_1 - 3$. It can be seen that eqn (5.21) does a better job than the Mooney-Rivlin equation in accounting for the results at higher strains; however, the fits at lower strains are poor. This failing is especially apparent in a conventional plot of engineering stress versus strain (**Figure 6**), in which the low strain data is not suppressed.

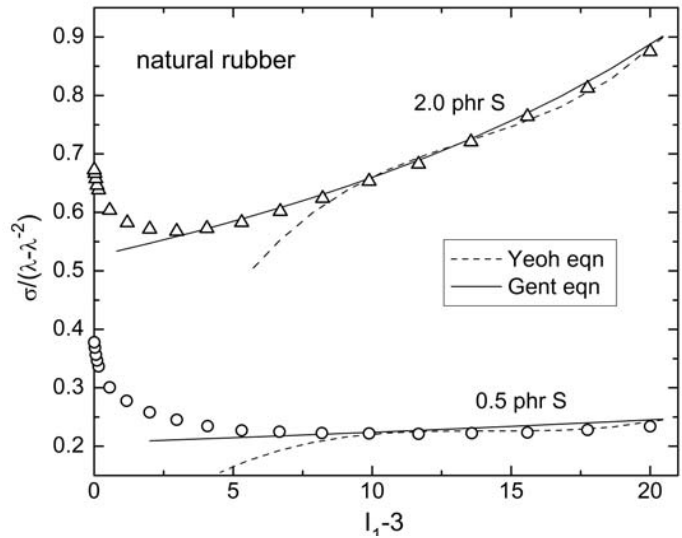

Figure 4. Data from Figure 3 plotted versus $I_1$-3. The dashed line is the fit of eqn (5.22) and the solid line the fit of the eqn (5.26).

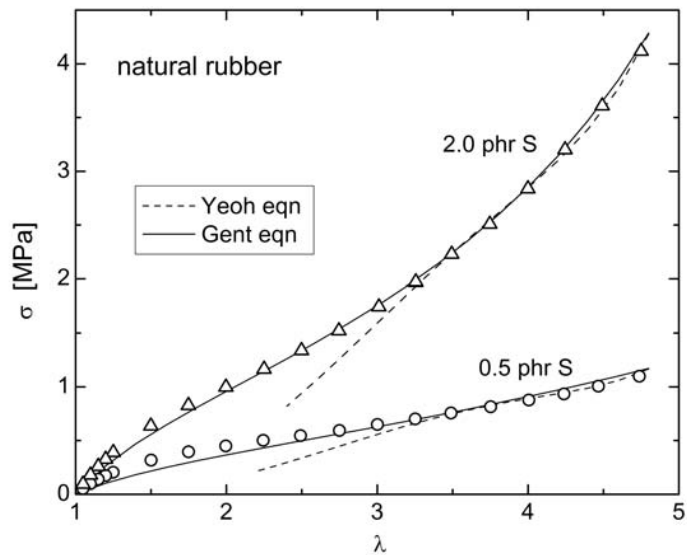

Figure 5. Data from Figure 3 plotted as the engineering stress versus extension ratio. The dashed line is the fit to eqn (5.22) and the solid line the fit to eqn. (5.26).

Gent [24] proposed a simple, empirical equation, in which the strain energy similarly depends only on the first strain invariant according to

$$W = -C_{10} \left( I_m - 3 \right) \ln \left( 1 - \frac{I_1 - 3}{I_m - 3} \right)$$

(5.24)

However, the deformation measure $I_1$ is limited to a maximum value, $I_m$, representing the effect of finite chain extensibility

$$I_1 \leq I_m$$

(5.25)

For uniaxial strain the engineering stress in the Gent model is

$$\sigma = 2 C_{10} \left( 1 - \frac{I_1 - 3}{I_m - 3} \right)^{-1} \left( \lambda - \lambda^{-2} \right)$$

(5.26)

It can be seen in Figure 3, Figure 5, and Figure 6 that eqn (5.26) is quite accurate at moderate to high strains. Even the deviation at low strains is small, given the limitation of only two adjustable parameters. In these figures, $I_m = 128$ and 52 for the NR networks crosslinked with 0.5 and 2.0 phr[§] sulfur, respectively. Although in practice it functions as a fitting parameter, from its definition $I_m$ must be related inversely to the shear modulus, since both depend (oppositely) on the network chain length [25,26]. Variations on this idea of introducing limits in the constitutive equation to account for finite chain extensibility have been proposed [25,27,28,29,30], including relaxing the incompressibility requirement [31]; however, these lack the economy of eqn (5.24).

## 4.2. Chain models

As discussed in Chapter 1, the force to distend a long (*ca.* 100 backbone bonds), flexible chain is directly proportional to the displacement, until the gauche conformers become depleted with approach to full extension. The end-to-end distance of the chain has a Gaussian distribution, which is unmodified by intermolecular interactions as long as the latter do not depend on chain configuration. (An exception would be liquid crystalline polymers, in which intermolecular forces have a strong dependence on chain orientation.) Calculation of the stress-strain relationship is derived classically from two extremes of how a chain can be envisioned to respond to deformation, the affine [6,32,33] and phantom [6,34,35] models. Both ignore any explicit consideration of interactions between chains including their uncrossability; the difference between the two models is how individual chains react to a macroscopic deformation. In the affine model the network junctions displace in proportion to the applied strain; that is, the ratio of the chain length before and after straining is the same as the ratio of the macroscopic dimensions. This corresponds to dominant interchain interactions that are independent of chain configuration. The force for a single chain was given by eqn (1.26) in Chapter 1

$$f_e = \frac{3k_B T}{\langle r^2 \rangle_0} r \tag{5.27}$$

where the subscript zero signifies the undeformed state. The strain energy is

$$W_{chain} = \frac{3k_B T}{2 \langle r^2 \rangle_0} \left( r^2 - \langle r^2 \rangle_0 \right) \tag{5.28}$$

---

[§] The concentration of an ingredient in a rubber formulation is usually expressed as "parts per hundred rubber" (phr), which is its weight divided by the weight of the polymer. The polymer content varies from 90% (rubber bands and gum tubing) to less than 20% (pencil erasers) of the mass of the compound.

Since the chains react independently, the total energy is obtained by summing over all $N$ chains per unit volume

$$W_{affine} = \frac{3N k_B T}{2} \left( \frac{\langle r^2 \rangle}{\langle r^2 \rangle_0} - 1 \right) \tag{5.29}$$

where $\langle r^2 \rangle = \sum r^2 / N_c$. In rectilinear coordinates

$$\langle r^2 \rangle = \langle x^2 \rangle + \langle y^2 \rangle + \langle z^2 \rangle \tag{5.30}$$

Since the network is isotropic, for an affine displacement

$$\frac{\langle r^2 \rangle}{\langle r^2 \rangle_0} = \left( \langle \lambda_1^2 \rangle + \langle \lambda_2^2 \rangle + \langle \lambda_3^2 \rangle \right) / 3 \tag{5.31}$$

and the strain energy is

$$W_{affine} = \frac{N k_B T}{2} \left( I_1 - 3 \right) \tag{5.32}$$

This result is identical to eqn (5.8) with the elastic constant $C_{10}$ identified as the number density of network chains times their thermal energy. For shear strains

$$\sigma_{affine} = N k_B T \gamma \tag{5.33}$$

so that $N k_B T$ equals the shear modulus. Assuming incompressibility, the engineering stress for uniaxial strain is

$$\sigma_{affine} = N k_B T \left( \lambda - \lambda^{-2} \right) \tag{5.34}$$

The alternative to the affine model is to consider Gaussian chains that interact only through their crosslinks, or network junctions. The chain segments can pass through one another and occupy the same volume simultaneously (that is, there is no excluded volume); hence, the network strands are referred to as phantom chains. While junctions at the surface of a phantom network deform affinely, all other crosslink sites fluctuate around their average position. This configurational freedom reduces the entropy, lowering the elastic modulus from the affine prediction. The strain energy for the phantom network is

$$W_{phantom} = \left[ 1 - 2/\widehat{f} \right] \frac{N k_B T}{2} \left( I_1 - 3 \right) \tag{5.35}$$

Both classical models, and indeed all molecular theories of rubber elasticity, predict the strain energy and the elastic modulus to be directly proportional to the number of network chains, $N$, and thus to the crosslink density. The phantom chain equations differ from eqns (5.32) to (5.34) of the affine model only by the factor in brackets containing the junction functionality, $\widehat{f}$. For bonding of just one pair of chains, the crosslink junctions are tetrafunctional, and the strain energy and elastic stresses for a phantom network are one-half that of an affine network having the equivalent number of crosslinks.

Elasticity models only consider chains that contribute to the mechanical stress, neglecting "elastically inactive" chains, such as those that are dangling, form a loop, or have $\widehat{f} = 2$. This leads to the use of cycle rank, $\widehat{R}$, to characterize the network topology [36]. The cycle rank, or number of independent cyclic paths, is equal to one-half the number of elastically effective network chains, $N_{eff}$. Eqn (5.35) for the strain energy of phantom network can be rewritten

$$W_{phantom} = \frac{\widehat{R} k_B T}{2} \left( I_1 - 3 \right) \tag{5.36}$$

For the common case of $\widehat{f} = 4$

$$N_{eff} = N(1 - 2M_c/M_n) \tag{5.37}$$

where $M_n$ refers to the number average molecular weight prior to crosslinking and $M_c$ is the molecular weight between crosslinks

$$M_c = \frac{\rho}{N} \tag{5.38}$$

Eqn (5.37) has the same form as eqn (1.13) from Chapter 1, describing the effect of chain ends on the glass transition temperature. Absent chain ends ($M_n \gg M_c$), the number density of network junctions is given by

$$\mu = \frac{2N}{\phi} \tag{5.39}$$

In common with phenomenological approaches to network elasticity, the classical models suffer from neglect of the limited deformability of the chains. This can be accounted for by invoking non-Gaussian statistics to describe the chains at higher strains. For a single chain this means extending the series expansion of the inverse Langevin function, eqn. (1.16) in Chapter 1, beyond the first term. This function accounts for the reduction in configurations, and thus entropy, of the chain as it becomes highly stretched [6,37]. For a network of chains, however, the problem becomes enormously complex, since the strain energy depends on the stretch, orientation, and spatial distribution of each chain. One approach is to develop a full-network model, which is then solved either for selected modes of deformation [38,39] or only approximately [40]. More progress has been made by representing the bulk elastomer as a simplified network of a few chains deforming in an assumed fashion relative to the bulk strain. Examples include the 3-chain [41], 4-chain [33,42], and the 8-chain models [43]. The latter, known as the Arruda-Boyce model, employs the inverse Langevin function, with the strain energy expressed as a series in the first strain invariant

$$W = G\left[\frac{1}{2}(I_1 - 3) + \frac{1}{20\hat{N}}(I_1^2 - 9) + \frac{11}{1050\hat{N}^2}(I_1^3 - 27) + \frac{19}{7000\hat{N}^3}(I_1^4 - 81) + \frac{519}{673750\hat{N}^4}(I_1^5 - 243) + ...\right]$$

(5.40)

Here $G$ is the shear modulus and the parameter $\hat{N}$ is related to the finite extensibility of the chains. The engineering stress for this model is

$$\sigma = 2G\left(\lambda - \lambda^{-2}\right)\left[\frac{1}{2} + \frac{1}{10\hat{N}}(I_1) + \frac{33}{1050\hat{N}^2}(I_1^2) + \frac{76}{7000\hat{N}^3}(I_1^3) + \frac{2595}{673750\hat{N}^4}(I_1^4) + ...\right] \quad (5.41)$$

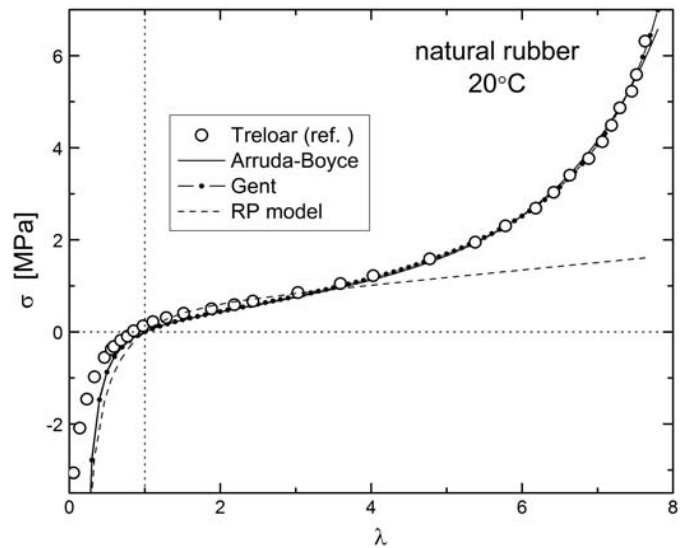

Figure 6. Engineering stresses for two NR networks in compression and tension [44]. The solid curve is the fit of eqn (5.41) with $G = 0.23$ MPa and $\hat{N} = 19.6$ and the dash-dotted line is the fit of eqn (5.26) with $C_{10} = 0.12$ MPa and $I_m = 83$. Also shown (dashed line) is the fit of eqn (5.57) to the low strain tensile data, which gives poor accuracy for high extensions and in compression.

In **Figure 7** the stress data of Treloar [44] for a vulcanized NR measured in uniaxial tension and compression are shown, along with the fit of eqn (5.40) to the tension data. Although the tension data are well described to high strains, using the same fit values of $G$ and $\hat{N}$ gives a poor description of the compression results. This is a failing molecular models have in common with phenomenological models.

There is a significant amount of literature comparing various elasticity models to experimental results [20,45,46,47,48,49,50,51]. An evaluation of twenty different approaches in terms of their ability to represent high strain data encompassing a range of deformation modes found the phenomenological Gent model (eqn (5.24)) [24] and the Arruda-Boyce chain model [43] to be the best among those with only two adjustable parameters [46]. These two models give very similar results [52], as shown in **Figure 7**.

### 4.3.    Constraint models

The cross-link sites in a real network are embedded in a high concentration of neighboring chain segments. At typical cross-link densities, spatially neighboring junctions are not topological neighbors; that is, the volume existing between a directly connected pair of junctions will contain many other junctions. The dense packing of chains (sometimes referred to as entanglements, although these local constraints are not necessarily manifested equivalently to the long-range topological interactions discussed in chapter 3) imposes limitations on the segment fluctuations. Neutron spin echo studies provide direct evidence that crosslink motions in networks are less than phantom network predictions [53,54]. The effect of these intermolecular interactions are ignored in the phantom chain model (even the uncrossability of chains), but overestimated in the affine model. A more accurate treatment of the mutual effect of neighboring chains has to consider the manner in which the interactions modify the stress in a strain-dependent manner. Constraint models of rubber elasticity have the general form

$$\sigma = \sigma_{phantom} \left[ 1 + f_{cnstr}(\lambda) \right] \qquad (5.42)$$

in which the function $f_{cnstr}$ reflects steric hindrances to Brownian motion of the network. From eqn (5.35)

$$\sigma_{phantom} = \left[ 1 - 2/\hat{f} \right] N k_B T \gamma \qquad (5.43)$$

for shear strains, and

$$\sigma_{phantom} = \left[ 1 - 2/\hat{f} \right] N k_B T \left( \lambda - \lambda^{-2} \right) \qquad (5.44)$$

for uniaxial strains.

The original constraint model of Flory [55] assumed the effect of constraints was to restrict fluctuations specifically of the network junctions. These constraints operate within a spatial domain, which deforms affinely with the macroscopic strain. The engineering stress for uniaxial strain in the constrained-junction (CJ) model is given by [55]

$$\sigma_{CJ} = \sigma_{phantom} \left[ 1 + \frac{2}{\bar{f} - 2} \frac{\lambda K(\lambda^2) - \lambda^{-2} K(\lambda^{-1})}{\lambda - \lambda^{-2}} \right] \tag{5.45}$$

where incompressibility is assumed. The function $K(\lambda^2)$ is defined in terms of functions of the strain

$$K(x) = \frac{B_1 \dot{B}_1}{B_1 + 1} + \frac{B_2 \dot{B}_2}{B_2 + 1} \tag{5.46}$$

where

$$B_1(x) = \frac{\kappa (x^2 - 1)}{(x^2 + \kappa)^2} \tag{5.47}$$

$$\dot{B}_1(x) = \frac{\partial B_1}{\partial x^2} = B_1 \left[ \frac{1}{x - 1} - \frac{2}{x + \kappa} \right] \tag{5.48}$$

$$B_2(x) = \frac{x^2}{\kappa} B_1(x^2) \tag{5.49}$$

and

$$\dot{B}_2(x) = \frac{\partial B_2}{\partial x^2} = \frac{1}{\kappa} \left[ x^2 \dot{B}_1 + B_1 \right] \tag{5.50}$$

The parameter $\kappa$ is a measure of the severity of the constraints. Eqn (5.45) reduces to the phantom and affine models respectively for $\kappa = 0$ and $\infty$.

Subsequently, Flory and Erman [56] extended the CJ model, introducing an additional parameter, $\xi$ ($\geq 0$), to allow for departures from uniformly affine distortion of the domain of constraints. A large value of $\xi$ implies that the constraints are alleviated by extension of the network more rapidly than if the domains deformed affinely. In this extended constrained junction (ECJ) model, $K(\lambda^2)$ is still given by eqn (5.46), but the functions $B_1(\lambda^2)$ and $B_2(\lambda^2)$ become

$$B_1(x) = \frac{\kappa^2 \left[ x^2 - 1 - x^2 \xi (x - 1) \right]}{\left[ x^2 + \kappa + \kappa x^2 (x - 1) \right]^2} \tag{5.51}$$

and

$$B_2(x^2) = x^2 \left[ \kappa^{-1} + \xi (x - 1) \right] B_1 \tag{5.52}$$

with the corresponding derivatives for $\dot{B}_1$ and $\dot{B}_2$. Substitution of these quantities into eqn (5.45) yields the engineering stress for the ECJ model, with the original CJ model recovered for $\xi = 0$. In **Figure 8** the ECJ model is compared to elastic mechanical results for NR at two different crosslink densities [57]. The model is fit to the tension data and, as seen, deviates strongly from the compression results.

In both the CJ and ECJ models, the constraints exert their effect directly on the network junctions. Alternatively, the chain's center of gravity can be the locus of the constraints [58,59], or more generally, the constraints can act all along the entire chain, as in the continuously-constrained chain (CC) model [60]. The uniaxial engineering stress for the CC model is

$$\sigma_{cc} = \sigma_{phantom}\left[1 + \frac{\hat{f}}{\hat{f}-2}\int_0^1 \Omega(z)\frac{\lambda K(\lambda^2)-\lambda^{-2}K(\lambda^{-1})}{\lambda - \lambda^{-2}}dz\right] \qquad (5.53)$$

where $K(\lambda^2)$ is again given by eqn (5.46). $\Omega(z)$ is a distribution function describing the effectiveness of the constraints at each point along the chain ($\Omega(z) = 1$ if the strength of the constraints is uniform) and $z$, which varies from 0 to 1, is the relative distance of a point on the chain from the junction site. So in the CC model the constraint parameter in eqns (5.47) – (5.50) is not constant, but varies along the chain according to

$$\kappa(z) = \overline{\kappa}\left[1 + \frac{(\hat{f}-2)^2 z(1-z)}{\hat{f}-1}\right] \qquad (5.54)$$

where $\overline{\kappa}$ is a constant, adjusted to fit experimental data. As seen in **Figure 8**, the CC model is slightly better than the ECJ model in describing the NR data; however, the problem remains that both tension and compression results cannot be simultaneously fit with good accuracy.

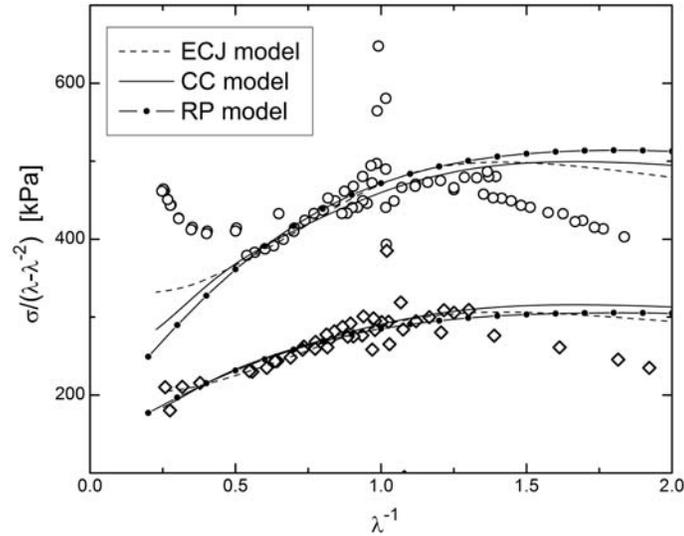

Figure 7. Reduced stresses for two NR networks in compression and tension [57]. The dashed curves are the fit of eqns (5.45), (5.51), and (5.52), the solid line the fits of eqn (5.53), and the dashed-dotted line fits to eqn (5.57), all to the tensile data. Note that experimental inaccuracies are magnified by plotting the ratio of the engineering stress and the strain function $\lambda - \lambda^{-2}$.

An idea of long-standing attraction is that the elastic response of a rubbery network reflects the same entanglement interactions that give rise to the plateau in the dynamic modulus of high molecular weight polymers (Chapter 3). There have been various attempts to connect the $C_{01}$ term of the Mooney-Rivlin equation ((5.10)) to the concentration of entanglements. Graessley [61] included this effect by expressing the stress as the sum of contributions from the phantom network and from the entanglements, with the latter proportional to the plateau modulus of the polymer in its precursor melt state

$$\sigma = \sigma_{phantom} + G_N f_N \tag{5.55}$$

The "trapping factor" $f_N$ represents the fraction of entanglements with paths covalently bonded to the network [62]. This expression does not address the non-linearity of the mechanical behavior.

Rubinstein and Panyukov [63,64] also proposed identifying the constraints with the chain entanglements, however with $f_N = 1$. Depending on the assumptions made about how these entanglements restrict fluctuations of the network chains and how this varies with strain, the engineering stress in the RP model is either [63]

$$\sigma = \sigma_{phantom} + G_N \frac{\lambda - \lambda^{-2}}{\lambda - \lambda^{1/2} + 1} \tag{5.56}$$

or [64]

$$\sigma = \sigma_{phantom} + G_N \frac{\lambda - \lambda^{-2}}{0.74\lambda + 0.61\lambda^{-1/2} - 0.35} \tag{5.57}$$

Although in principle $G_N$ is the plateau modulus of the uncrosslinked polymer, in practice it serves as an adjustable parameter. As shown in **Figure 8** the RP gives results quite close to the CC model, and thus shares its failure to describe simultaneously results from more than one mode of deformation. And since they are based on modifications of the phantom chain approach, the constraint models do not account for the upturn in the stress at high strains, as illustrated in **Figure 7**.

Table 2. Network chain densities from **Figure 8**.

| Determination | N (mole/m³) | |
|---|---|---|
| | 1.0 phr dicumyl peroxide | 2.0 phr dicumyl peroxide |
| crosslinking chemistry* | 66 | 136 |
| ECJ model (eqns (5.45), (5.51), and (5.52)) | 162 | 264 |
| CC model (eqn (5.53)) | 130 | 206 |
| RP model (eqn (5.57)) | 218 | 264 |

* assuming complete reaction of the peroxide crosslinker with $\hat{f} = 4$ [57].

In Table 2 are collected the results from **Figure 8**, comparing the number of network strands deduced from the cross-linking reaction to the value of $N$ obtained by application of the various models. The latter are all substantially higher, indicating that $\hat{f}$ may be greater than the value of four used in the calculation. The usual assumption is that peroxide crosslinking of natural rubber yields a stable tertiary radical that terminates by combination to form tetrafunctional junctions [65].

Eqns. (5.56) and (5.57) embody ideas drawn from slip-link models of rubber elasticity [66,67,68]. Heinrich et al. [69,70] developed a related model that introduced constraint release effects that, as discussed in Chapter 1 describing the rheology of entangled polymer liquids, reduce the severity of the entanglement constraints, yielding more phantom-like behavior.

### 4.4. Role of molecular motions in the elastic response

The picture that emerges from rubber elasticity theories is one of chains frustrated by intermolecular constraints in their effort to achieve all the configurations available to an isolated chain; thus, the elastic response reflects the effect the network structure exerts on the chain dynamics. Since the topology and motion of chain molecules are both governed by the same intramolecular and intermolecular potentials and correlations, there have been various efforts to analyze rubber elasticity from study of the chain dynamics [71,72,73,74,75,76,77,78]. The underlying assumption is the usual one of FDT – equilibrium fluctuations of the network segments and junctions are related to the linear relaxation behavior (see Chapter 1).

The coupling model of relaxation [2] addresses the effect on the local dynamics of the same constraints restricting the network configurations. According to this model, at short times before unbalanced forces and torques have attained sufficient magnitude, the chains exhibit exponential time dependence

$$C_{ph}(t) \sim \exp[-(t/\tau_{ph})] \quad , \quad t < t_c \tag{5.58}$$

with $\tau_{ph}$ the junction relaxation time. (Eqn (5.58) applies at all times for phantom chains, since they experience no intermolecular constraints.) After some temperature-insensitive time scale, $t_c$, the average relaxation rate of the segments becomes slowed down by intermolecular cooperativity. From experimental data for amorphous polymers [23-26,27,28], $t_c$ is found to have a magnitude between $10^{-12}$ and $10^{-11}$ s. For times longer than $t_c$, which fall within the usual, experimentally accessible realm, the dynamics of the network chains are slowed by their interactions with neighboring chains, causing the simple exponential decay to become stretched (*cf.* eqn 1.7 in Chapter 1)

$$C_{cnstr}(t) = [-(t/\tau_\alpha)^{\beta_K}] \quad , \quad t > t_c \tag{5.59}$$

In the model the important metric of intermolecular constraints is the coupling parameter, $n = 1 - \beta_K$, so that the Kolhrausch exponent varies inversely with the strength of these interactions. At $t = t_c$ eqns (5.58) and (5.59) can be set equal to one another, whereby the correlation time at longer times is given by

$$\tau_\alpha = \left[ \beta_K t_c^{\beta_K - 1} \tau_0 \right]^{1/\beta_K} = \left[ (1-n) t_c^{-n} \tau_0 \right]^{1/(1-n)} \qquad (5.60)$$

This relation (*cf.* eqn 2.17 in Chapter 2) makes clear the strong, nonlinear slowing down of the dynamics at longer time, since $\tau_\alpha \gg \tau_0$. In the model the onset of intermolecularly constrained relaxation is accompanied by an increase of the activation energy, $E_a$, which from eqn (5.60) is related to $E_a^0$, the microscopic conformational energy barrier to segment motion, as

$$E_a = E_a^0 / (1-n) \qquad (5.61)$$

$E_a$ is the value prevailing at longer times when constraints are dominant, although a consequence of these constraints is deviation from Arrhenius behavior

$$\tau_\alpha = \tau_\infty \exp\left[ -E_a / RT \right] \qquad (5.62)$$

This means that the observed $E_a$ varies with temperature. (Note that the inverse correlation of fragility and the Kohlrausch $\beta_K$ described in Chapter 2 follows from eqn (5.61)). Eqn (5.61) is the same relation that leads to the coupling model prediction of a breakdown in time-temperature superpositioning in the glass transition zone, due to the stronger coupling (larger $n$) of the segmental modes in comparison to the global chain modes (Chapter 6). There is an analogous relationship to eqn (5.60) between the respective pre-exponential factors, $\tau_{0,\infty}$ and $\tau_\infty$, for uncoupled and coupled relaxation.

Table 3. Comparison of constraint models of elasticity and coupling model of relaxation

| Network alteration | effect on constraints | $\kappa^*$ | $\beta_K$ | $E_a^*/E_a$ | $\tau_\infty^*/\tau_\infty$ |
|---|---|---|---|---|---|
| larger $N$ | smaller | smaller | larger | smaller | smaller |
| larger $\widehat{f}$ | larger | larger | smaller | larger | larger |
| larger $\lambda$ | smaller | --- | larger | smaller | smaller |
| diluent | smaller | smaller | larger | smaller | smaller |

*eqns (5.45) - (5.54)

The expectation is that the severity of the constraints on the junctions, giving rise to departures of the elastic response from phantom-like behavior, is evident directly in the junction dynamics. This means that experimental variables that affect the parameter $\kappa$ in the constraint models (which characterizes the domain of the constraints on the fluctuations) will similarly affect the coupling parameter $n$ describing intermolecular cooperativity of the junction dynamics (see Table 3). This is illustrated with NMR spin-lattice relaxation measurements [79] on PTHF networks with various molecular weights between crosslinks.

The network junctions were tris(4-isocyanatophenyl) thiophosphate, so that the NMR experiment measured the motion of the junctions. From spin-lattice relaxation times, spectral density functions corresponding to the Fourier transforms of the *C(t)* in eqn (5.59) were obtained. The obtained coupling parameters are displayed in **Figure 9**, showing that the junctions in the more crosslinked PTHF experience stronger intermolecular constraints (larger *n*) [77,79]. A similar trend in the parameter $\kappa$ of the constraint elasticity models is observed [80].

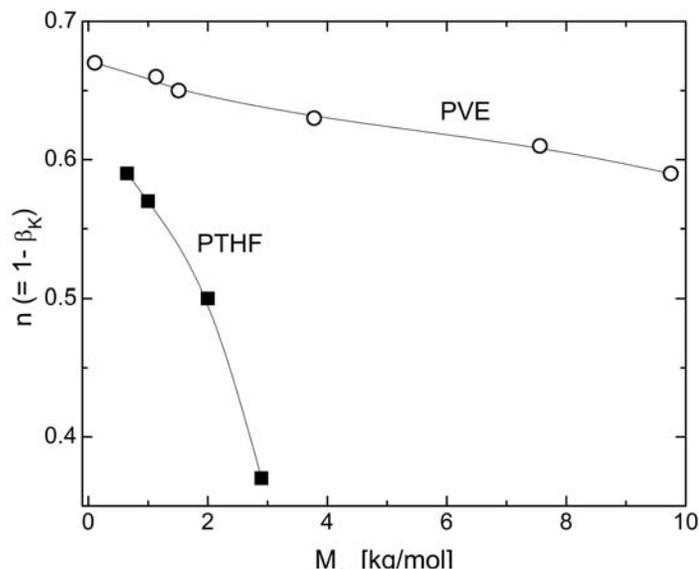

Figure 8. Coupling parameter *versus* molecular weight between crosslink junctions for polytetrahydrofuran (solid symbols) [77,79] and polyvinylenthylene (open symbols) [78] networks.

As predicted by eqn (5.61), the apparent activation energy increases with *n* (**Figure 10**). The product (1-*n*)$E_a$ is constant, equal to the activation energy in the absence of cooperative dynamics, $E_a^0$ = 26 kJ/mol. This value is of the same magnitude as the conformational energy barrier of the polyether backbone. It is unaffected by intermolecular interactions, reflecting only the intramolecular barriers to changes in rotameric state, as well as the static, mean-field background potential. Also in **Figure 10** are data representing the most crosslinked network with added diluent. The motion of small molecules is very rapid on the time scale of the chain dynamics, so the diluent alleviates the intersegmental constraints, decreasing both the coupling parameter and $E_a^*$.

The motion of the junctions followed by the NMR experiment is, of course, coupled to the motion of the chains connecting the junctions. The differences between the CJ and CC models are artifacts of the respective assumptions of these approaches. Thus, the dynamics of networks measured by a technique such as dielectric or mechanical spectroscopy, which reflects the motion of all segments rather than specifically the junctions, can be analyzed in a similar manner. However, in a randomly crosslinked polymer the chain segments have different mobilities depending on their proximity to a junction. Segments closer (topologically or spatially) to a crosslink will have their motions more retarded, especially for large junction functionalities, wherein the confluence of many strands imposes more severe steric constraints. For this reason the segmental relaxation dispersion (peak in the dielectric or mechanical loss) will be inhomogeneously broadened, whereby its shape cannot be interpreted directly in terms of intermolecular cooperativity (*viz.*, eqn (5.59) with $n = 1 - \beta_K$).

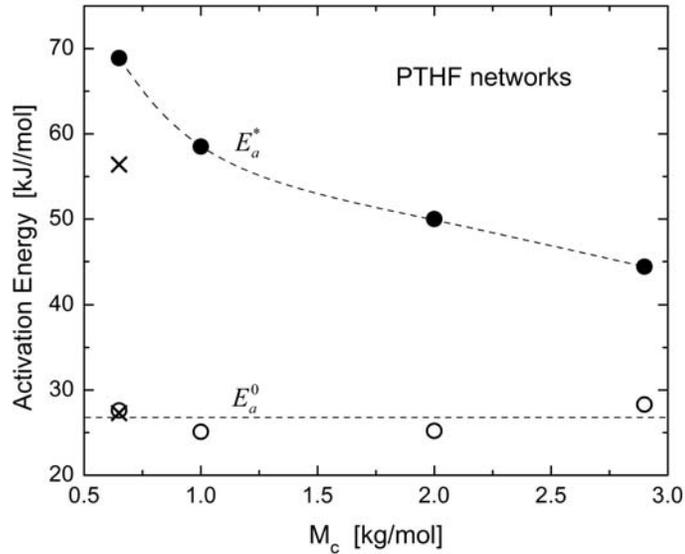

Figure 9. Apparent activation energy (filled symbols) measured by [31]P NMR [79] and non-cooperative activation from eqn (5.61) (hollow symbols) using the coupling parameters in Figure 9 [77], as a function of the molecular weight between crosslinks. The crosses are data for the diluted network.

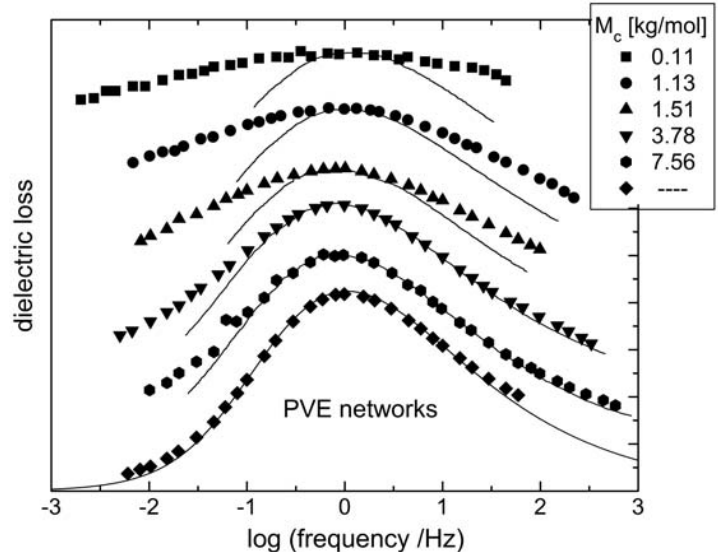

Figure 10. Peak in the dielectric loss for PVE networks of increasing crosslinking from bottom (uncrosslinked) to top; curves have been vertically displaced for clarity. The solid lines are calculated using the coupling parameters determined from applying eqn (5.61) to the fragility curves in Figure 12. Data from ref. [78].

This broadening is evident in dielectric spectra for a series of polyvinylethylene (PVE or 1,2-polybutadiene) networks (**Figure 11**), which have large $\hat{f}$ (due to the free radical crosslinking). The peaks deviate strongly from the stretched exponential form, and the coupling parameter cannot be obtained directly from fitting the spectra to eqn. (5.59). However, since $E_a^0$ must be the same for each PVE sample (since it represents the intrinsic barrier to conformational transitions of the isolated chain), $n$ can be determined as the value that satisfies eqn (5.61). (In molecular dynamic simulations this heterogeneous broadening is circumvented, since the dynamics of the chain end-to-end vector can be determined directly. Simulation results for rubber networks confirm that both the coupling parameter and the relaxation time increase systematically with increasing degree of crosslinking [81].)

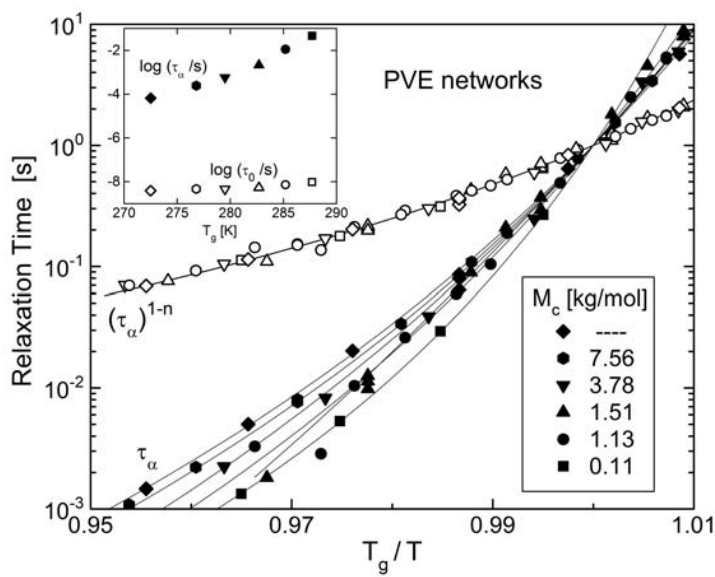

Figure 11. $T_g$–normalized Arrhenius plots of the dielectric relaxation times as measured (solid symbols) and after raising to the power of $\beta_K$ (open symbols; see eqn (5.60)). The inset displays the relaxation times for each network (identified by $T_g$) at $T = T_g$. Data from ref. [78].

Determining the coupling parameter from $E_a$ is opposite to the procedure used to analyze the NMR data on PTHF networks. Since $\tau_\alpha(T)$ is non-Arrhenius, rather than the apparent activation energy *per se*, the fragility, or $T_g$-normalized activation energy (Chapter 2), is used. The results are shown in **Figure 12**. After removal of the effect of intermolecular cooperativity by assuming a value for the coupling parameter, the calculated activation energy and relaxation times for segmental relaxation are essentially independent of crosslink density. Using the values deduced for $n$, eqn (5.59) is plotted in **Figure 11**. The experimental peak is much broader than the deduced Kohlrausch function because of the inhomogeneous broadening arising from the distribution of environments in the random crosslinked PVE.

The assumption that the non-cooperative activation energy, $E_a^0$, is independent of crosslink density can be evaluated directly from dielectric relaxation measurements on the PVE networks [82]. The Johari-Goldstein secondary peak is present in the spectra, with the obtained JG relaxation times showing an Arrhenius temperature-dependence with an activation energy $E_a^{JG}$ (**Figure 13**). According to the coupling model, these $\tau_{JG}$ can be indentified with $\tau_0$ in eqn (5.60); thus, $E_a^0 = E_a^{JG}$. From the parallelism of the slopes in **Figure 13,** it is seen that the measured $E_a^{JG}$ are indeed independent of crosslinking and equal to the value for linear PVE. However, the value of $E_a^{JG} = 44$ kJ/mol [82] is much larger than that deduced from the analysis in **Figure 12**. The discrepancy comes from the assumption that $\tau_0$ is constant. As seen in **Figure 13**, $\tau_{JG}$ (= $\tau_0$) becomes smaller with crosslinking, opposite to the effect on $\tau_\alpha$. This is an interesting anomaly, seen also during physical aging of glassy PVE [83].

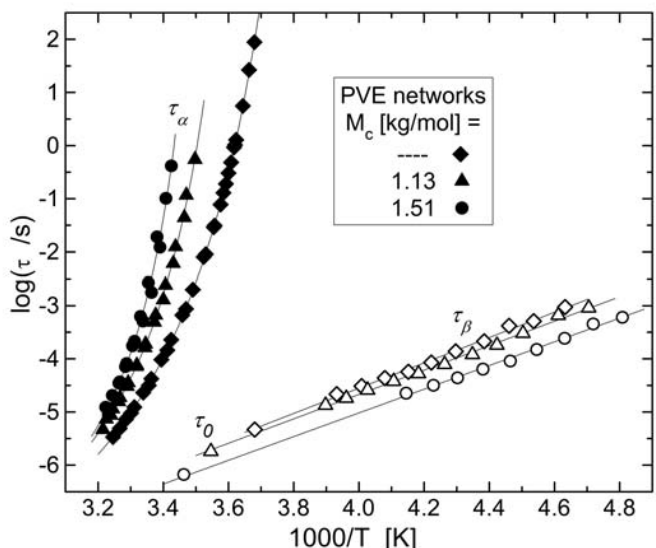

Figure 12. Relaxation times for the local segmental (filled symbols) and JG secondary (open symbols) processes in linear and crosslinked PVE. The open symbols with dotted center denote the value of $\tau_0$, obtained by extrapolating $\tau_\beta(T)$ to the temperature at which for $\tau_\alpha = 3.5 \times 10^4$ s for each sample. Data from ref. [82].

## 4.5. Alternative network structures

In developing an elastomeric compound, the particular polymer selected for the formulation has minimal effect on the network properties *per se*. That is, while the chemical structure of a rubber affects various aspects of performance (e.g., chemical and heat resistance, friction, adhesion, gas permeability, biocompatibility, etc.), the mechanical behavior up through moderate deformations is essentially the same for all flexible-chain polymers. Even the fracture and fatigue properties, in the absence of strain-induced crystallization, are not especially dependent on the polymer. Of course, the chemical structure has an indirect effect, through its influence on the crosslinking reaction, but circumventing the usual compromise between stiffness and strength requires unconventional network architectures. One method is to incorporate ionic crosslinks, which dissociate under stress. Elastomeric "ionomers" incorporate an ionizable monomer or have an inorganic salt grafted pendant to the main chain. The result is a crosslink density that effectively decreases with the extent of deformation, as the junctions progressively dissociate during large deformation to yield substantial toughness (**Figure 2**) [3,84,85]. However, this same mechanical lability confers large unrecovered strain (permanent set) and susceptibility to degradation. In these respects, ionically crosslinked rubbers are similar to vulcanizates with polysulfidic bonds, although an additional drawback the former is poor processability, due to the limited control over when the ionic bonds form.

Several approaches to improving the mechanical properties of elastomers by alternative network structures are discussed below.

**4.5.1 Interpenetrating polymer networks.** An interpenetrating polymer network (IPN) refers to a blend ostensibly consisting of co-continuous, interlocking networks (catenanes) of the constituents [86,87,88]. This co-continuity is achieved by kinetic retention of an initially homogenous mixture of small molecules during subsequent polymerization and crosslinking, the formation of a network precluding phase segregation of the immiscible polymers [89,90]. The objective is a micro-heterogeneous morphology stabilized by physical interlocking; however, segmental interpenetration of the two network structures may not actually be obtained, so that in practice the designation IPN connotes the method of preparation, rather that the actual blend morphology.

IPNs can be made simultaneously or sequentially. A simultaneous IPN is formed from the polymerization and crosslinking of premixed monomers or linear prepolymers. During network formation, the tendency for phase separation is promoted by the increasing molar mass of the constituents (which reduces the combinatorial entropy) but opposed by their concomitantly slower dynamics. In a sequential IPN [91,92,93], one network is swollen in monomers of the other component, and the latter is then rapidly polymerized and crosslinked, often using radiation curing. Phase separation is more extensive for sequential IPNs, but co-continuity can still be achieved. "Semi-IPN" is used in reference to an IPN in which one component remains uncrosslinked [94,95]. The $T_g$ of an IPN is expected to be intermediate to the neat component $T_g$'s, as shown in **Figure 14** for an IPN [96] and **Figure 15** for a semi-IPN [97]. Even though the constituents are incompatible, small phase domains are obtained, homogenizing the properties. The glass transition temperature becomes size-dependent for domain sizes on the order of or less than the cooperativity length scale, so that a change in the component $T_g$ does not necessarily indicate mixing on the segmental level. A distribution of phase sizes gives rise to a very broad glass transition dispersion, and for this reason IPNs have attracted interest as materials for acoustic damping and vibration isolation [98,99]. Judging the degree of homogeneity and interpenetration of an IPN from its glass transition behavior can be ambiguous when the component $T_g$'s are close.

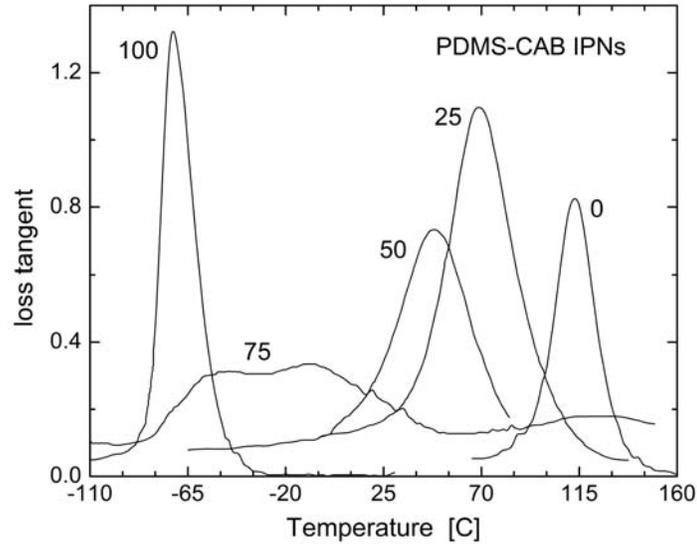

Figure 13. Loss tangent for polydimethylsiloxane – acetate butyrate cellulose IPNs having the indicated weight % of PDMS. Data from ref. [96]

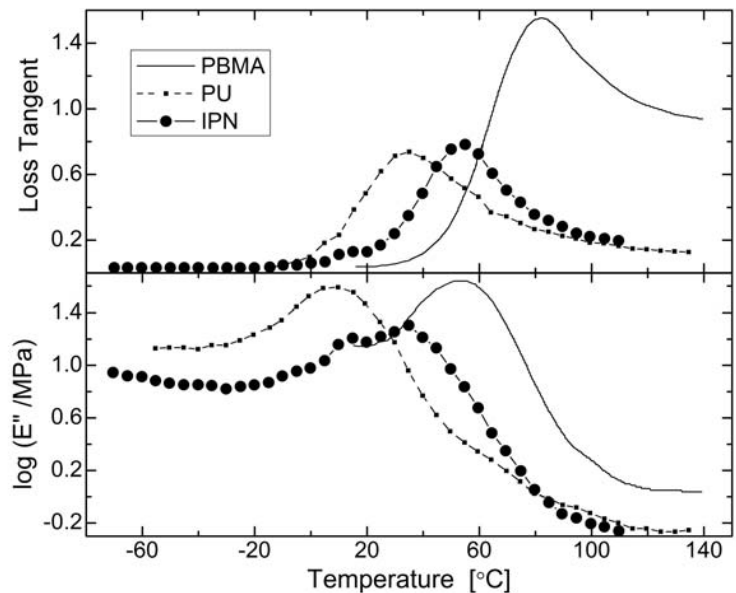

Figure 14. Loss tangent for linear polybutylmethacrylate, polyurethane network, and their semi- IPN. Data from ref. [97].

The premise of IPN formation is the absence of interference among the polymerization and cross-linking reactions, although such interferences can be unavoidable. Grafting between the components may occur, with consequences for both the morphology (such as the inhibition of phase separation) and properties [86,100,101,102,103]. While from the structure of IPN's it is expected that mechanical performance will be additive with regard to the component properties, interlocked networks can confer greater mechanical integrity than a completely phase separated blend morphology. This potential for better properties is illustrated in Figure 16 [102] for an IPN of a fluoroelastomer and nitrile rubber blend, which exhibits a maximum in failure properties as a function of composition.

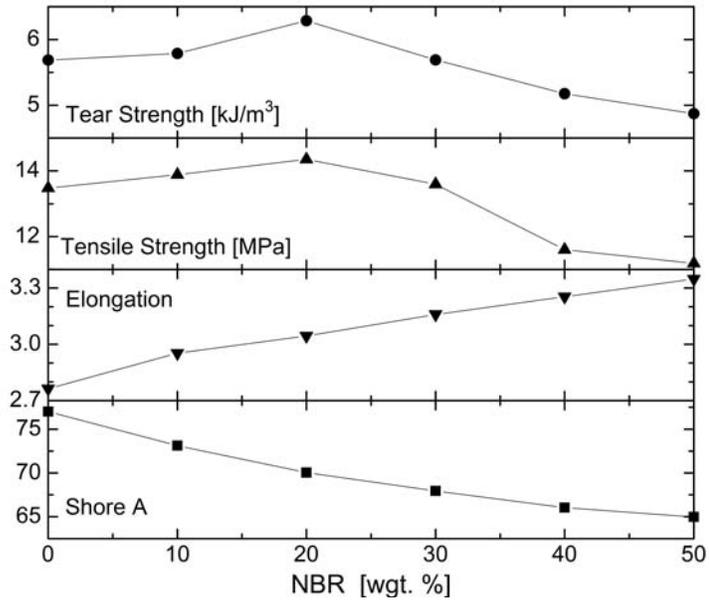

Figure 15. Properties of an IPN based on nitrile rubber and poly(vinylidene fluoride-co-hexafluoroproylene). With increasing NBR content, the material becomes more compliant; however, the tensile and tear strengths exhibit a maximum. Data from ref. [103].

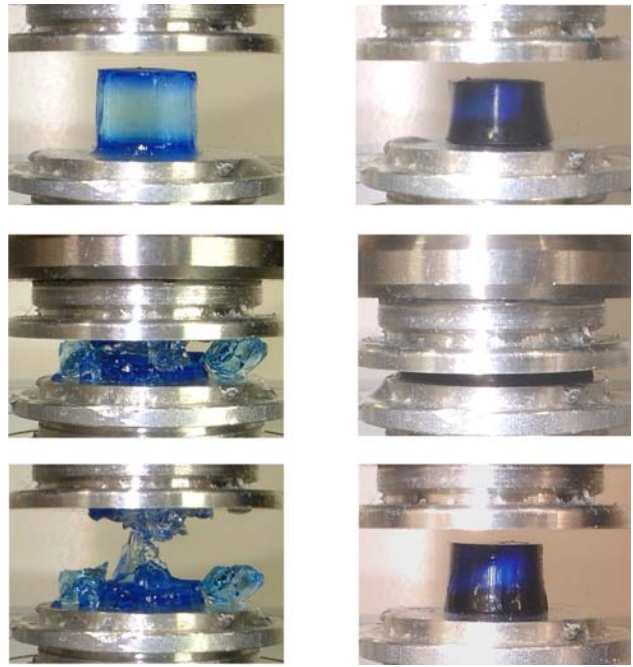

Figure 16. Compression of (a) a poly(2-acrylamido-2-methylpropanesulfonic acid) hydrogel and (b) an IPN hydrogel based on poly(2-acrylamido-2-methylpropanesulfonic acid) and polyacrylamide, showing the greater resistance to fracture of the IPN. From ref. [115] with permission.

Phase homogeneity in IPNs can be greatly enhanced if a co-solvent is used; for example, a popular type of IPN are hydrogels. These are dilute aqueous IPNs, which react to changes in pH, temperature, or the presence of specific chemicals. Their biocompatibility, high water content, and ability to respond to stimuli have led to use of IPN hydrogels as biomaterials for tissue engineering [104,105,106] and for controlled drug delivery [107,108,109,110]. These materials, sometime confusingly referred to as double-network hydrogels, can exhibit enhanced performance [111,112,113,114]. For example, Figure 17 [115] shows the remarkable increase in compressive strength of an IPN hydrogel compared to the corresponding single network hydrogel; however, the components of this material have very different crosslink densities, an essential element in the improvement of properties (see sections 4.5.3 and 4.5.4).

**4.5.2 Double networks.** A special type of IPN is a double network rubber, in which the same chain segments belong to the two networks, both of which are oriented. A double network is formed by two crosslinking processes, the second while the elastomer is in a deformed state (usually tension). Since the properties of an elastomeric network depend not only on the crosslink density, but also on the distribution and orientation of the chains, the properties of double networks can be quite different from those of the corresponding single network at equal degrees of crosslinking. Double networks are a means to increase both the stiffness and strength of rubber. They can arise spontaneously via chain scission [116,117], strain-induced crystallization[118,119], due to the presence of reinforcing fillers [120], or as described in Chapter 8 by crosslinking liquid crystalline elastomers while in their ordered state. The effect of aging on stress-relaxation ("chemical stress relaxation") [121] and of trapped entanglements on the elastic modulus [122,123] both involve the contribution of inadvertent double networks.

Analysis of double networks relies on the independent network hypothesis [124,125,126,127], which assumes that the behavior of the component networks are uncoupled, so that the mechanical response is the sum of the individual contributions. The equilibrium configuration of each network corresponds to that existing during network formation, so the strain energy of a double network is

$$W_{DN}(\lambda) = W_1(\lambda) + W_2(\lambda/\lambda_X) \qquad (5.63)$$

in which the subscripts refer to the first and second network, and $\lambda_X$ is the stretch ratio during the second crosslinking (the initial network being formed in the unstressed, isotropic state). The strain is referenced to the undeformed network; hence, $\lambda = \lambda_1$ and $\lambda_2 = \lambda/\lambda_X$. The residual strain at zero stress of the double network represents the state for which the forces from the initial network, which is subsequently extended, are balanced by those from the second network, which is compressed. Although from eqn (5.63) the stress of a deformed double network is just the sum of the stresses from the component networks, modeling of the mechanical behavior is limited by the poor ability of elasticity theories to simultaneously describe tensile and compressive strains (**Figure 7** and **Figure 8**).

The orientation of the network chains gives rise to anisotropic mechanical properties [128,129,130,131,132,133]. Generally the modulus increases parallel to the curing strain, with negligible changes in the transverse direction. This modulus enhancement is illustrated in **Figure 18** [134] for a series of elastomers having the same total crosslink density, but varying $\lambda X$. For double networks reinforced with carbon black, lower electrical conductivity [135] and suppression of the Payne effect (Chapter 5) [136] have been observed, both consistent with less filler aggregation due to the strain imposed during curing.

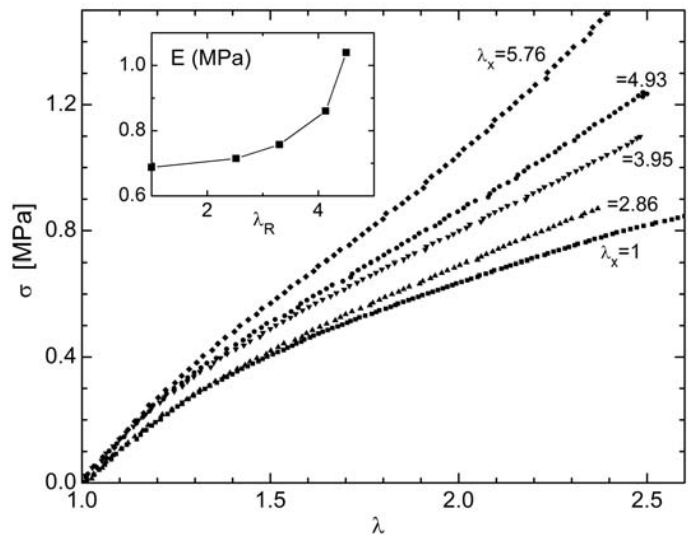

Figure 17. Engineering stress as a function of stretch ratio for double networks having the indicated strain during the second crosslinking. The crosslink density was fixed at 102 mol/m$^3$. The inset shows the stress at 100% strain. Data from ref. [134].

Although the experimental results are scattered [128,129,130,133], a double network structure affects the strength of the elastomer only weakly. Given their higher modulus, this means that double networks can be used to circumvent the usual compromise between stiffness and strength. Moreover, double networks of strain-crystallizing rubbers having substantially larger fatigue lifetimes (**Figure 19** [129,137]). This appears to result from retention of crystallinity through the minimum of the strain cycle [137], analogous to the greater fatigue life of crystallizing rubbers subjected to non-relaxing strain cycles [138].

As illustrated in **Figure 20**, double networks are birefringent in the absence of external stress [132,134], contrary to the stress optical law (eqn. 5.14 in Chapter 5). This stress-free birefringence is consistent with constraint models of rubber elasticity, which predict that the ratio of the birefringence to the true stress varies with strain. For example, the CC model gives [139]

$$\Delta n_{cc} = \Delta n_{phantom} \left[ 1 + \frac{\widehat{f}}{\widehat{f} - 2} \int_0^1 \Omega(z) \frac{B_1(\lambda) - B_1(\lambda_\perp) + b_{cc} \left( B_2(\lambda) - B2(\lambda_\perp) \right)}{\lambda^2 - \lambda^{-1}} dz \right] \qquad (5.64)$$

where $\Delta n_{phantom}$ is a constant proportional to the true stress. The parameter $b_{cc}$, which has a value between 0 and 1, accounts for the contribution of the constraints to the birefringence. Since the effect of constraints is strain-dependent, the apparent stress optical coefficient varies with strain, as confirmed in measurements on silicone [140] and polybutadiene [141] rubbers. Thus, even if the stress and birefringence of the components networks are both additive quantities, double networks will exhibit non-zero $\Delta n$ in the stress-free state [134].

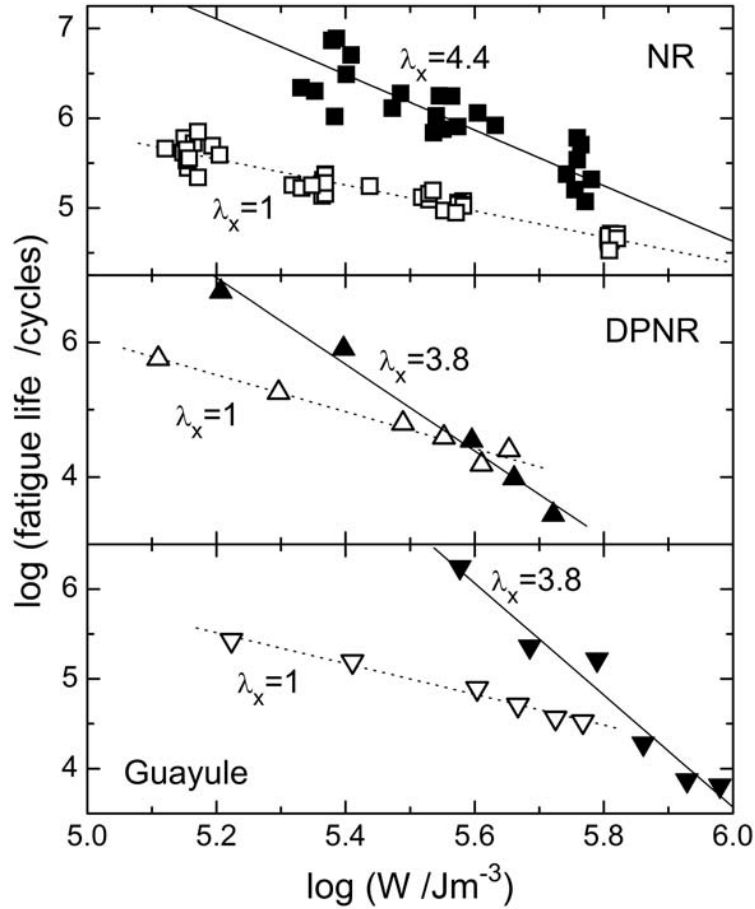

Figure 18. Double logarithmic plots of the mechanical fatigue life *versus* recoverable strain energy for single (hollow symbols) and double (solid symbols) networks of natural rubber (squares), deproteinized natural rubber (triangles), and guayule rubber (inverted triangles). Measurements were for uniaxial extension parallel to the curing strain. Data from ref. [129, 137].

**4.5.3 Bimodal networks.** A bimodal network is an elastomer in which a portion of the chains between crosslink junctions are very short and the remaining are very long. Model bimodal networks are prepared by end-linking a mixture of low and high $M_w$ precursor chains [142]. Scattering experiments show the inhomogeneity of bimodal networks on the size scale of the network mesh [143], with some indication in PDMS networks of segregation of the short and long chains [144]. Clustering of the short and long chains has also been seen in simulations of bimodal networks [145]. However, the respective orientations of the long and short chains are not greatly different, as found in infrared dichroism measurements on deuterium-labeled networks of poly(tetrahydrofuran) (PTHF, also known as polytetramethylene oxide) (**Figure 21**) [146]. In bimodal networks with a preponderance of short chains, the mesh size of the network was found to be essentially the same as for unimodal networks of the same short chains [147]. This is not the case for bimodal networks composed primarily of long chains, in which this correlation length relates to the entanglement spacing [147].

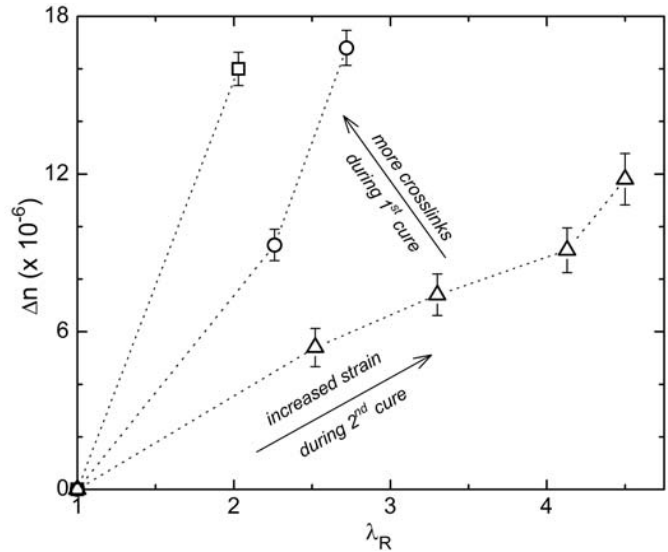

Figure 1. Birefringence measured at zero stress for various double networks, identified according to their residual stretch ratio, all having the same total crosslink density (=102 mol/m³). The latter, and hence $\Delta n$, increases with increase in either the proportion of crosslinks in the first network or the strain during formation of the second network. Data from ref. [134].

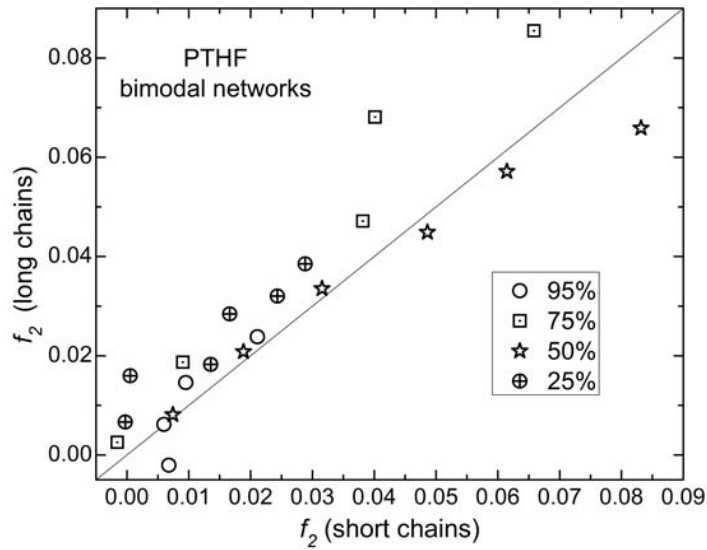

Figure 20. Hermans orientation function (see eqn 4.18) measured for PTHF bimodal networks having the indicated mole percent of short chains. $M_n = 11,076$ and 1,560 g/mol for the long and short chains, respectively. The solid line corresponds to equal orientation. Data from ref. [146].

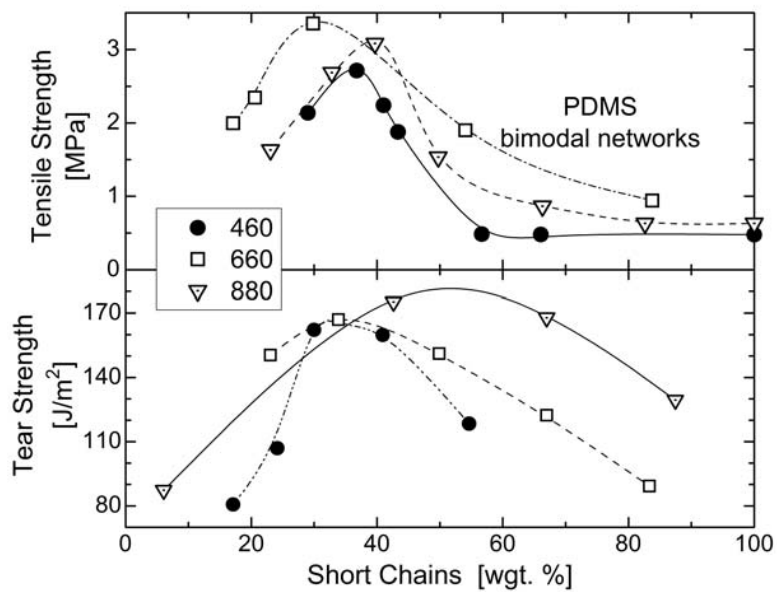

Figure 21. Failure properties of PDMS bimodal networks with short chains having the indicated $M_n$; for the long chains $M_n = 21,300$ g/mol. Data from ref. [148,149].

The appeal of bimodal networks is that they can exhibit substantially higher toughness and strength [148,149] (**Figure 22**). Such results are contrary to the early ideas of rubber elasticity, according to which failure is due to rupture of the shortest network chains [150], implying the short chains in a bimodal network would weaken the elastomer. The improved properties require that the short chains be low in molecular weight (typically a few hundred g/mol) and lower by as much as a factor of 100 than the molecular weight of the longer chains [151]. The data in **Figure 22** exhibit maxima, indicating that the number density of the short chains should be large, typically around 95 mol%, which still corresponds to a low weight fraction. Note that the converse – a small weight fraction of long chains – also improves toughness, although such materials are not elastomeric [152,153]. The improved properties of bimodal networks are ascribed to a synergy between the high modulus of the short chains and the extensibility of the longer ones [142,151]. Experiments on polyethylene oxide [154] and polytetramethylene oxide [155] indicate a greater propensity to strain crystallize, presumably because the short chains orient sooner to serve as nucleating sites. This enhanced strain-crystallization would contribute to better failure properties.

**4.5.4 Heterogeneous miscible networks.** The components of an IPN are immiscible, with crosslinking used to minimize phase segregation; however, the morphology of IPNs is rarely homogeneous on the segmental level. Hydrogels employ water as a solvent to compatibilize the components and, as described above, enhanced properties can be achieved [111,112,113,114,115]. If the polymers comprising a network are thermodynamically miscible, there is no driving force for phase separation and a composition can be achieved that is uniform on the sub-nanometer level. Various commercial materials are based on miscible polymer blends, although a homogeneous phase morphology *per se* does not necessarily provide improved mechanical properties. The expectation is that the blend properties will be some average of those of the pure components. Thus, there is a second important requirement - a large disparity in the respective crosslink densities of the components. This disparity causes their respective contributions to the network mechanical response to differ diametrically, in this manner emulating bimodal networks and the hydrogel in (**Figure 17**), but without the need for end-linking or a solvent. Unlike IPNs, the components are uniformly mixed and there is only a one network. Chains of the more crosslinked component become highly stretched upon deformation of a heterogeneous miscible network, raising the stress locally, while chains of the lightly crosslinked component are able to alleviate over-stressing and consequent spread of rupture nuclei, which would otherwise lead to macroscopic failure.

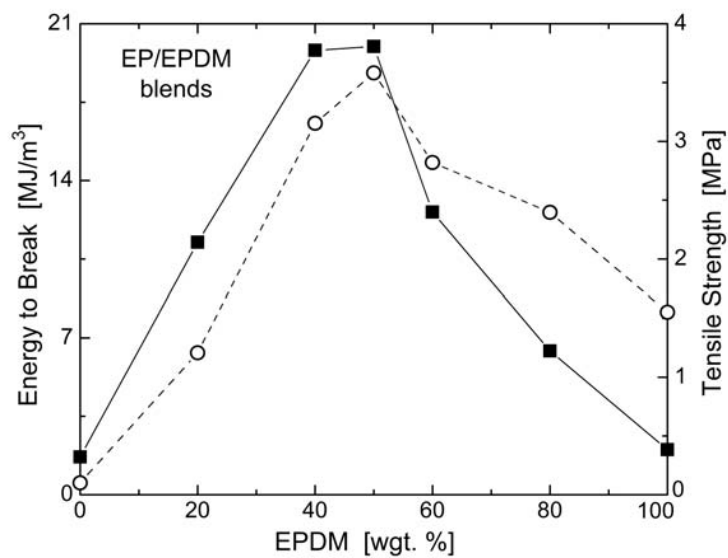

Figure 22. Fracture energy (filled squares) and stress at break (open circles for vulcanized blends of EPDM with ethylene-propylene copolymer, in which the curative level was adjusted to maintain a constant crosslink density of the EPDM (Young's modulus = 1.5 ± 0.1 independent of composition). Data from ref. [156].

There are two approaches to achieve a miscible mixture of a highly crosslinked polymer with one that is lightly or even uncrosslinked. One means is to use chemically identical components, except that only one has (a few mole percent) crosslink sites. Examples would be EPDM mixed with ethylene-propylene copolymer and butyl rubber mixed with polyisobutylene.

Some data illustrating the first method is shown in **Figure 23** for EPDM (a random terpolymer of ethylene, propylene and a diene[**]) blended with a random copolymer of ethylene and propylene (EP) [156]. The blend is miscible over significant ranges of backbone composition [157]. The EP is unreactive to sulfur vulcanization, so a large disparity in crosslink density can be achieved. At constant modulus the failure properties of EPDM/EP networks exhibit a maximum at approximately equal concentration of the components.

The second approach to heterogeneous networks is blending chemically different polymers that are thermodynamically miscible and have different crosslinking reactivities. One example is the mixtures of PI and PVE. Miscibility of natural rubber and high vinyl polybutadiene was first reported by Bartenev and Kongarov [158], and subsequently investigated by various groups [159,160,161]. Thermodynamic miscibility is the result of a fortuitous near-equivalence of the polarizabilities (and thus van der Waals energies) of the repeat units for the two polymers [162,163]. Although infrared spectroscopy indicates an absence of specific interactions [164], the interaction parameter is negative, implying a lower critical solution temperature (LCST) [162]. The blend is attractive for hetergeneous networks because tertiary vinyl carbons do not sulfurize [65], so that only PI can be vulcanized.

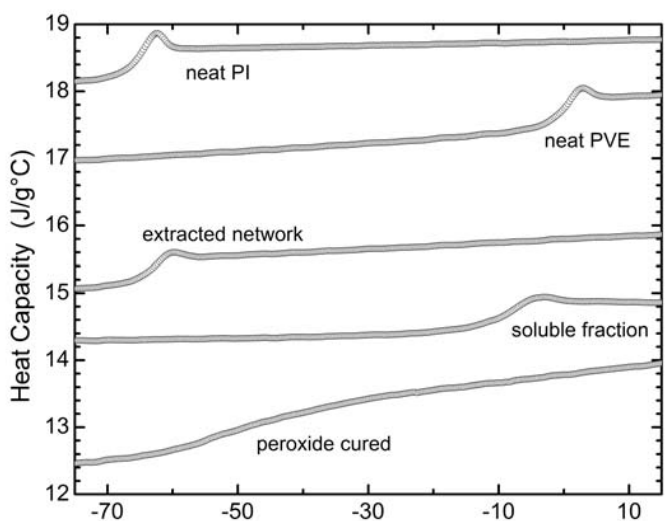

Figure 23. Fracture energy (filled squares) and stress at break (open circles for vulcanized blends of EPDM with ethylene-propylene copolymer, in which the curative level was adjusted to maintain a constant crosslink density of the EPDM (Young's modulus = $1.5 \pm 0.1$ independent of composition). Data from ref. [156].

This selectivity of sulfur crosslinking in PI/PVE blends is illustrated in DSC curves (**Figure 24**) [165]. The gel phase is essentially PI, while the soluble fraction is the unreacted PVE. However, the chemical changes effected by the crosslinking reaction reduce the miscibility of PI with the PVE, so that some degree of heterogeneity of the phase morphology arises, governed by the rate of curing [165]. Alternatively, organic peroxide can be used to crosslink the blend. Although the PVE is more reactive than PI to free radicals, the method does not give a sufficient disparity in the respective degrees of cure. Consequently, the mechanical properties are poorer, notwithstanding the more homogeneous phase morphology (**Figure 25**) [165]. Similar to the behavior of bimodal networks, the best mechanical properties correspond to the largest disparity in crosslinking.

[**] EPDM rubbers employ ethylidene norbornene, dicyclopentadiene, 1,4-hexadiene, vinyl norbornene, or norbornadiene for the unsaturated repeat unit at concentrations up to 12%.

Another miscible blend with components having different crosslinking chemistries is 1,4-polychloroprene (PC) and epoxidized 1,4-polyisoprene (EPI). The miscibility arises from specific interaction of the chlorine with the oxirane group [166,167]. While EPI sulfur vulcanization is similar to that of polyisoprene, PC can be crosslinked using zinc and magnesium oxides without an organic accelerator. The zinc oxide produces ether crosslinks through the tertiary allylic chorine, and the magnesium oxide extends the reaction by scavenging the chlorine ion biproducts [168]. The distinct crosslink mechanisms of PC and EPI afford control of the rate and degree of network formation in their blends. It has been shown that a miscible morphology is retained after curing, although the properties obtained for the blend were only intermediate between those of the pure components (**Figure 26**) [169].

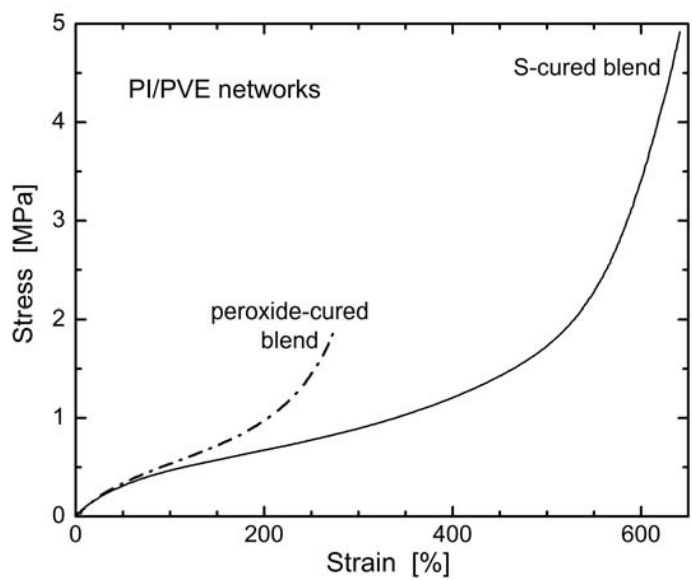

Figure 24. Engineering stress versus tensile strain for the sulfur and peroxide cured blends of equal weights of PI and PVE. The PVE is relatively unreactive to the S-vulcanization, with the consequent disparity crosslink density of the components yielding better mechanical properties. Date from ref. [165].

### 4.5.5 Deswollen networks.

The stretch ratio of the chains comprising a network, $\lambda_0$, is related to the macroscopic extension ratio, $\lambda$, according to

$$\lambda_0 = \left( \frac{V}{V_x} \right)^{1/3} \lambda \qquad (5.65)$$

where $V_x$ and $V$ are the volumes during crosslinking and subsequent deformation, respectively. Eqn (5.65) indicates that deswelling of a network prepared by crosslinking diluted chains results in a compacted or "supercoiled" network [170]. This compaction of the chains is seen in radius of gyration measurements on deswollen ($V/V_x = 0.1$) polystyrene networks of different crosslink density (**Figure 27**) [171]. The behavior is analogous to networks swollen by a poor solvent, which similarly causes some collapse of the chains [172].

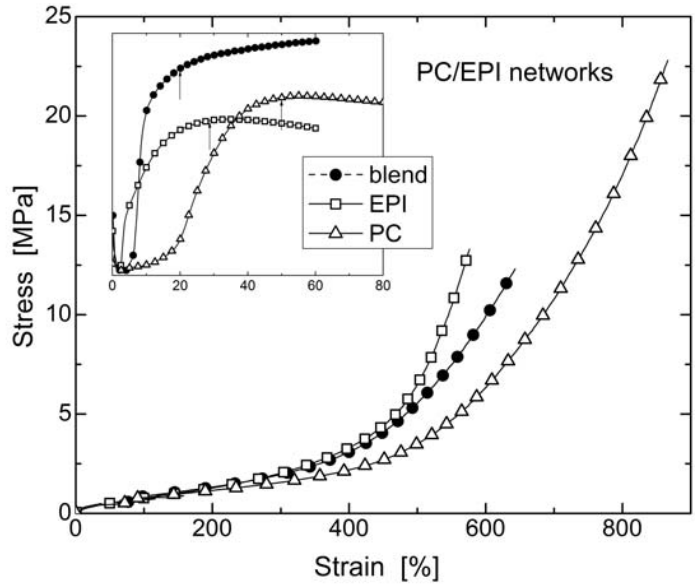

Figure 25. Engineering stress versus tensile strain for the equal weight blends of polychloroprene and 50 mole % epoxidized polyisoprene. The blend mechanical properties are intermediate to those of the pure components due to the near parity of the crosslink density of the components. The inset show the cure curves for the neat components and the blend, with the arrow indicating the state of cure at which the respective stress-strain curves were measured. Data from ref. [169].

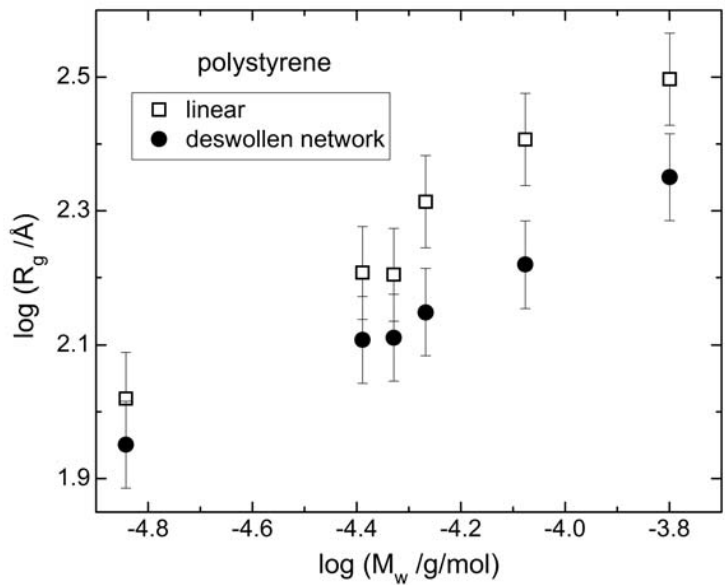

Figure 26. Radius of gyration of deuterated polystyrene chains dispersed in a melt (squares) and in a network crosslinked in a tenfold-swollen state and then dried (circles), as a function of the molecular weight of the respective melt or network chains. Data from ref. [171].

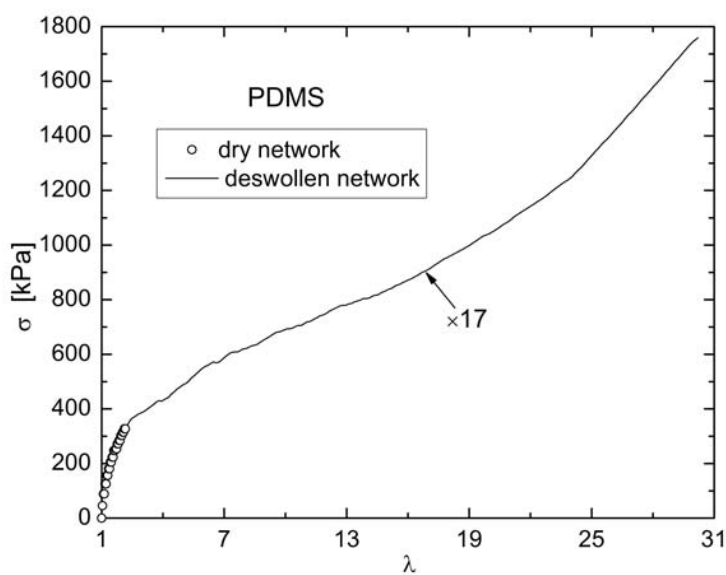

Figure 27. Engineering stress-strain curve for PDMS conventional network (circles) and formed in toluene solution at 25% volume fraction and tested after solvent removal (solid line). The data for the latter were multiplied by 17 to superpose the initial moduli. The stretch ratio at failure increases from 2 to about 30 for the deswollen elastomer. Data from ref. [175].

The highly compressed "internal state", in combination with a lowered degree of chain entanglements, causes deswollen networks to exhibit low modulus and very large failure strains [173,174]. This "super-elasticity" is illustrated in **Figure 28** comparing stress-strain curves for conventional and deswollen networks [175]. The greater packing of the collapsed chains increases the mass density (about 15% for $V_x \gg V$ in Vasiliev et al. [176]), which contributes to an increase in modulus. The result is a minimum in the dependence of the modulus on $V_x$ (**Figure 29**). The more compact chain configurations in deswollen networks also suppresses the rate of thermal (unoriented) cystallization [177].

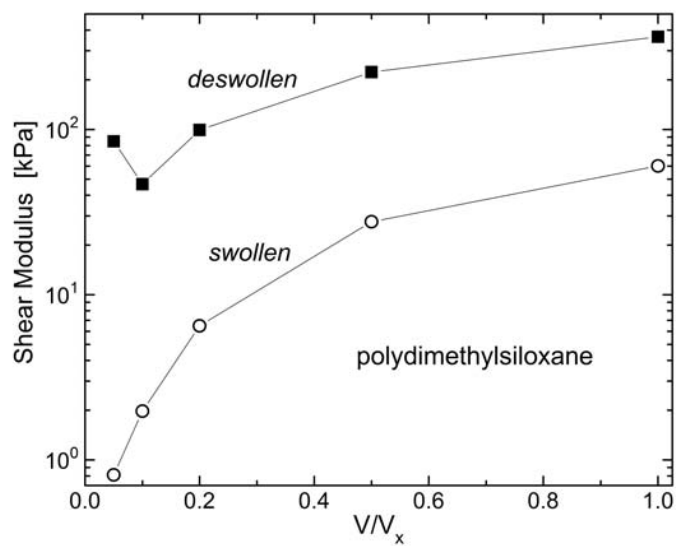

Figure 28. Shear modulus of PMDS crosslinked at the indicated volume fraction measured before (circles) and after (squares) deswelling (note that the ordinate scale is logarithmic). Data from ref. [176].